\documentclass[%
 reprint,
 amsmath,amssymb,
 aps, physrev,
]{revtex4-2}

\usepackage{graphicx}% Include figure files
\usepackage{dcolumn}% Align table columns on decimal point
\usepackage{bm}% bold math
\usepackage{hyperref}
\usepackage{ulem}
\usepackage{xcolor}

\newcommand{\meank}{\bar{k}}%\meank

\newcommand{\ddt}{\frac{d}{dt}}
\newcommand{\stateS}{\mathsf{S}}
\newcommand{\stateI}{\mathsf{I}}

\begin{document}

%\preprint{APS/123-QED}

\title{\textbf{Multi-strain spreading dynamics under arbitrary transmission kernels} }% 

\author{Sagar Kumar}
\affiliation{Network Science Institute, Boston, MA, USA}
\affiliation{Center for Health Informatics Program, Boston Children's Hospital, Boston, MA, USA}%Lines break automatically or can be forced with \\

\author{Moritz Laber}
\affiliation{Network Science Institute, Boston, MA, USA}
\affiliation{Complexity Science Hub Vienna, Vienna, Austria}

%\author{P\'eter L. Simon}
%\affiliation{Department of Applied Analysis and Computational Mathematics, E\"otv\"os Lor\'and University, Budapest, HU}

\author{Maimuna S. Majumder}
\affiliation{Center for Health Informatics Program, Boston Children's Hospital, Boston, MA, USA}
\affiliation{Department of Pediatrics, Harvard Medical School, Cambridge, MA, USA}

\author{Brooke Foucault Welles}
\affiliation{
 Network Science Institute, Northeastern University, Boston, MA, USA
}%
\affiliation{
 Department of Communication, Northeastern University, Boston, MA, USA
}%
\author{Istv\'an Zolt\'an Kiss}
\email{Contact author: istvan.kiss@nulondon.ac.uk}
\affiliation{%
 Network Science Institute, Northeastern University, London, UK}
 \affiliation{ Department of Mathematics, Northeastern University, Boston, MA 02115, USA
}%

\date{\today}

\begin{abstract}
Compartmental models of epidemic dynamics have long described the propagation of a single, immutable transmissible state through a population via pairwise contact, and multi-strain generalizations have extended this framework to incorporate mutation, competition, and cross-immunity. Here we study a minimal generalization with no sink states or feedback, in which transmission acts through an arbitrary column-stochastic kernel $Q$ on a finite set of strains, encoding mutation during transmission with no further structural assumptions. We derive the mean-field approximation for the well-mixed regime and show that it admits an exact closed-form solution for any $Q$, expressible as a single matrix exponential applied to the initial condition. A spectral decomposition of this solution reveals that the location of the long-time attractor and the rate of approach are governed by the eigenstructure of $Q$. We extend the analysis to structured populations via a pairwise mean-field approximation on regular contact networks, and validate both approximations against stochastic simulations. The framework provides an entry into the analysis of dynamical systems in which mutation and transmission occur on the same time scale, drawing parallels to the propagation of discrete signals through populations under noisy communication.
\end{abstract}


\maketitle

%\tableofcontents

\section{Introduction}

The classical compartmental models of epidemic dynamics, beginning with the Susceptible-Infected (SI) model and its extensions~\cite{kermack_contribution_1997, anderson_infectious_1991}, describe the propagation of a single transmissible state through a population via pairwise contact. Multi-strain generalizations have extended this framework in several directions. For example, models have been constructed for competing pathogens with cross-immunity~\cite{karrer_competing_2011, castillo-chavez_competitive_1996}, interactions between strains~\cite{funk_interacting_2010}, mutation during transmission via specified parametric structures~\cite{girvan_simple_2002}, and mutation along an explicit genotype network with strain-dependent immune interactions~\cite{williams_localization_2021}. Each of these models includes either some additional structure that constrains the kinds of strain-to-strain transitions the model permits, or includes some feedback or sink state to extend the dynamics beyond a simple spreading. While this additional dynamical machinery is essential for quantitative comparison with real disease spread, it can obscure more basic questions about how the long-time composition of a population depends on the rate and structure of mutation during transmission, and limits extensions of these frameworks to spreading processes beyond epidemiology (e.g., the spread of information, beliefs, or innovations through a population, where what is being transmitted is itself subject to distortion). 

In this work, we strip all of this back to study a minimal generalization of the Susceptible-Infected model in which transmission occurs through an arbitrary column-stochastic kernel, $Q$, acting on a finite set of strains, $\mathcal{A}$. Each transmission event from an individual infected with strain $j$ to a susceptible individual produces a new infection in strain $k$ with probability $Q_{kj}$. As such, the diagonal elements of $Q$ encode faithful transmission, and the off-diagonal elements encode mutation. No structure is imposed on $Q$ beyond column-stochasticity, so the model accommodates arbitrary mutation patterns within a single framework. This combination of generality and analytical tractability distinguishes the present work from prior multi-strain treatments, which have either fixed restrictive parametric forms for the mutation channel in order to retain closed-form analysis, or admitted more general mutation structures at the cost of analytical access to the long-time behavior. We refer to this as the ``Noisy'' SI model, due to the role played by the transition kernel $Q$, which is formally identical to a noisy channel in information theory~\cite{shannon_mathematical_1998, cover_elements_2006, mackay_information_2003}. This correspondence motivates a second reading of the model as approximating the propagation of a discrete signal through a population in which each transmission introduces fixed-rate distortion. Seen this way, the Noisy SI model can be considered as generalizing existing treatments of rumor and information dynamics~\cite{daley_epidemics_1964, maki_mathematical_1973, davis_phase_2020, rapoport_mathematical_1952} which have largely treated transmission either as faithful or as a single-state stochastic event. We develop this communicative interpretation in Sec.~\ref{sec:communication} such that the analytical results of the paper serve both as a contribution to the multi-strain epidemic literature and as a foundation for the communicative case study. The principal results of the paper concern the analytical tractability of this model. We derive a mean-field approximation for the well-mixed regime that admits an exact closed-form solution that is expressible as a single matrix exponential applied to the initial condition. Spectral decomposition of this solution reveals that the long-time attractor is governed by the eigenstructure of $Q$, with the leading eigenvector determining the location of the attractor in the space of strain densities, and the spectral gap controlling the rate at which trajectories converge to this distribution. We also develop a pairwise approximation for structured populations and find that network sparsity reshapes the long-time attractor in a manner qualitatively similar to increased channel noise. Both approximations are validated against stochastic simulations. In the communicative register, these spectral results translate directly to a statement about information loss: network sparsity, initial seed size, and the spectral structure of the channel together determine how rapidly source information is forgotten as the cascade propagates.

The remainder of the paper is organized as follows. Section~\ref{sec:model} introduces the stochastic model and its reaction-kinetic specification. Section~\ref{sec:wellmixed} develops the well-mixed mean-field approximation and its exact solution. Section~\ref{sec:structured} extends the analysis to networked populations via pairwise closure. Section~\ref{sec:spectral} characterizes the attractor through the spectral structure of $Q$. Section~\ref{sec:communication} develops the communicative reinterpretation and traces how the spectral structure governs information transmission through the population. We conclude with a discussion of regimes of applicability and connections to information-theoretic and cultural-evolution models in Section~\ref{sec:discussion}.

\section{Stochastic Model}\label{sec:model}
\begin{figure*}
    \centering
    \includegraphics[width=.9\textwidth]{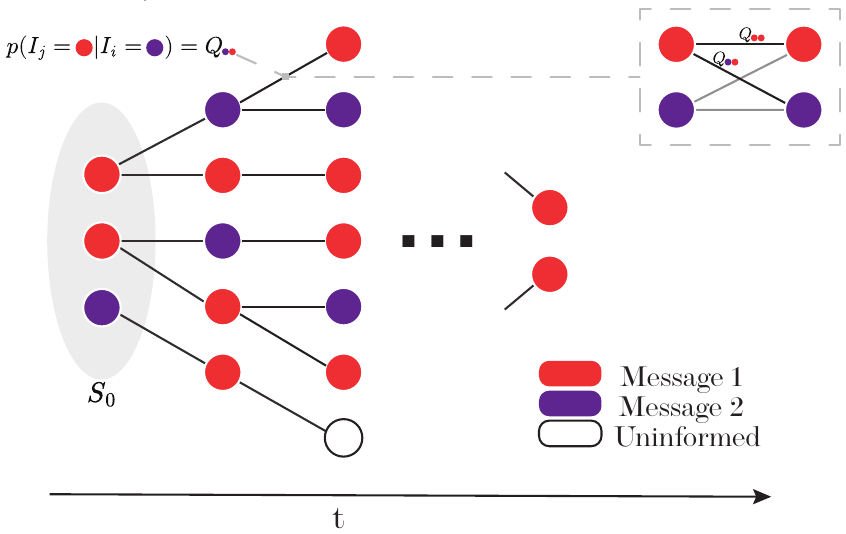}
    \caption{\textbf{Illustration of the Noisy SI model.} Each transmission event (edges going forward in time) is mediated by a noisy channel $Q$, such that the strain transmitted from an infected individual need not match the strain received by a newly infected individual, as indicated by the colors of the nodes. The probability that a susceptible individual infected by an individual in strain $j$ transitions into strain $k$ is $Q_{kj}$, the corresponding entry of the column-stochastic kernel $Q$. The diagonal of $Q$ encodes faithful transmission and the off-diagonal entries encode mutation.}
    \label{fig:model_sketch}
\end{figure*}

%\subsection{Susceptible-Infected Model} 

The Susceptible-Infected (SI) model is a special case in which the recovery rate of the Susceptible-Infected-Recovered (SIR) model of disease spread \cite{kermack_contribution_1997} is zero. As such, it partitions populations into two states: susceptible and infected. Its dynamics are defined by the single reaction 
\begin{equation}\label{eq:SI_reaction}
    \stateS + \stateI \xrightarrow{\beta} 2\stateI,
\end{equation} which describes an individual in the susceptible state coming into contact with an individual in the infected state and becoming infected with probability $\beta$.

To allow for noise or mutation in the spreading contagion, we now extend this so that instead of a singular infected state, there is a set of infected states, $\mathcal{A} = \{\stateI_1, \stateI_2, ..., \stateI_{|\mathcal{A}|}\}$. We assume that each transmission from an infected node to a susceptible node occurs over a noisy channel \cite{shannon_mathematical_1998}, $Q$, such that 
\begin{equation}\label{eq:general_reaction}
    \stateS + \stateI_j \xrightarrow{\beta Q_{kj}} \stateI_j+ \stateI_k
\end{equation}
where $Q_{kj}$ is the probability that a susceptible individual successfully infected by an individual in state $\stateI_j$ will transition into state $\stateI_k$. In a two-strain setting, the full set of reactions is thus 
\begin{align*}
    \stateS + \stateI_1 &\xrightarrow{\beta Q_{11}} 2\stateI_1 \\
    \stateS + \stateI_1 &\xrightarrow{\beta Q_{21}} \stateI_1 + \stateI_2 \\
    \stateS + \stateI_2 &\xrightarrow{\beta Q_{12}} \stateI_1 + \stateI_2 \\
    \stateS + \stateI_2 &\xrightarrow{\beta Q_{22}} 2\stateI_2.
\end{align*}

\section{Well-Mixed Mean Field}\label{sec:wellmixed}

\subsection{Deriving the Well-Mixed Noisy SI Mean Field}
Following the procedure outlined in \cite{kiss_mathematics_2017}, we can define a linear transition function
\[
f_{SI_k} = \beta \meank \sum_j^{|\mathcal{A}|} Q_{kj} [ SI_j ],
\]
which reflects the contribution from each strain to $I_k$, the number of individuals in state $\stateI_k$ at that time. Here, we use $\meank$ to refer to the average number of interactions per unit time, and $[ SI_j ]$ to denote the expected number of edges between susceptible individuals and individuals infected with strain $j$, which, for a well-mixed population, is simply the product of $I_j$ and the number of susceptible individuals, $S$. Because we do not allow transitions $\stateI_j \to \stateI_k$, there are no outgoing reactions, and this is the only transition function to be considered. Thus, the expected number of individuals in state $\stateI_k$ can be expressed as 

\[
    \frac{d}{dt}[ I_k ] = \beta \meank\sum_j^{|\mathcal{A}|} Q_{kj} [ SI_j ]. 
\]

We can express this as a single variable by introducing a state (column) vector $\bm{x}(t) = ([ I_1 ]/N, [ I_2 ]/N, ..., [ I_{|\mathcal{A}|} ]/N)$, given a population of size, $N$. For closed populations,
\begin{equation}\label{eq:closed_S}
    S = N - I_1 - I_2 -... -I_{|\mathcal{A}|}
\end{equation}
we can then write the mean-field approximation for noisy susceptible-infected dynamics in a homogeneous population as 
\begin{equation}\label{eq:NSI_mf}
    \frac{d}{dt}\bm{x}(t) = \beta \meank \left(1-\sum_k^{|\mathcal{A}|}\bm{x}_k(t)\right)Q\bm{x}(t).
\end{equation}

The form of~\eqref{eq:NSI_mf} is that of a logistic growth similar to the mean-field approximation for the standard Susceptible-Infected model, with the constants on the left setting a rate, the middle term being the fraction of the population that is still susceptible, and the last term being the fraction of the population in an infected state. Unlike the SI model, however, the well-mixed mean-field approximation of the Noisy SI model introduces $Q$ as ``mixing'' the infected states.

\subsection{Exact Solution to the Mean-Field Approximation}
This mean-field approximation is solvable for any $Q$ by introducing a time-dependent variable
\[
y(t)= 1-\sum_{k=1}^{|\mathcal{A}|} \bm{x}_k ,
\]
which translates to the fraction of the population still in the susceptible state. The system can then be written in the vectorial form as
\begin{equation}
\dot{\bm{x}}(t) = \beta \meank y(t) Q\bm{x}(t). 
\label{diffeq_vec}
\end{equation} 
We first derive a differential equation for $y$ by, in general, assuming that the sum of each column in $Q$ is the same. This is, assuming
\[
Q_j = \sum_{k} Q_{kj} 
\]
does not depend on $j$. As such, there is a number $c$, such that $c=Q_j$ for all $j$. For any column-stochastic matrix, $c = 1$ by definition, but for generality, we will allow it to vary. Now summing the differential equations of $\bm{x}_k$ for all $k$ yields
\begin{align*}
-\dot{y} &= \beta \meank  y \sum_{k=1}^{|\mathcal{A}|} \sum_{j=1}^{|\mathcal{A}|} Q_{kj}\bm{x}_j \\
& = \beta \meank  y \sum_{j=1}^{|\mathcal{A}|} \bm{x}_j \sum_{k=1}^{|\mathcal{A}|} Q_{kj} \\
&= \beta \meank  cy(1-y).
\end{align*}
This differential equation can be easily solved by separating the variables and decomposing the left side using partial fractions
\[
\frac{\dot y}{y-1} - \frac{\dot y}{y} = \beta \meank  c,
\]
which can be integrated as
\[
y(t)= \frac{y_0 e^{-\beta \meank ct}}{1-y_0 + y_0 e^{-\beta \meank ct}},
\]
where $y_0=y(0) \in (0,1)$ is the initial condition. 

Now we can consider our system~\eqref{diffeq_vec} as a linear system for the unknown vector $\bm{x}(t)$ with time-dependent coefficients. However, one can exploit the fact that the time-dependence is in an extremely special form, since each entry of the matrix $Q$ is multiplied by the same scalar $y(t)$. Hence the linear system can be solved in the same way as in the time-independent, i.e. autonomous case using the matrix exponential. Thus let us look for the solution of \eqref{diffeq_vec}  in the form
\[
\bm{x}(t) = \exp(z(t)Q) \bm{x}(0) .
\]
Substituting this form into \eqref{diffeq_vec} yields
\[
\dot{\bm{x}}(t) = \dot z(t) Q \exp(z(t)Q) \bm{x}(0)= \dot z(t) Q \bm{x}(t).
\]
Therefore we have $\dot z(t) = \beta \meank y(t)$ with the initial condition $z(0)=0$. Hence introducing 
\[
Y(t) = \int_{0}^{t} y(s) ds ,
\]
the solution of \eqref{diffeq_vec}  can be written as
\[
\bm{x}(t) = \exp(\beta \meank Y(t)Q) \bm{x}(0) .
\]
Integrating the function $y$ yields
\[
Y(t)=- \frac{1}{\beta \meank c} \ln \left( 1-y_0 + y_0 e^{-\beta \meank ct} \right) .
\]
Finally, we can express the limit $\bm{x}(\infty)=\lim_{t\to \infty} \bm{x}(t)$ explicitly in terms of the initial condition $\bm{x}(0)$ as follows 
\begin{equation}\label{eq:stable_mf}
    \bm{x}(\infty) =  \exp(\beta \meank Y(\infty)Q) \bm{x}(0),
\end{equation}
where 
\begin{equation}\label{eq:y_inf}
    Y(\infty)=- \frac{1}{\beta \meank c} \ln ( 1-y_0),
\end{equation}
with $1-y_0= \sum \bm{x}_i(0)$.

\subsection{Agreement with Simulation}

\begin{figure}
    \centering
    \includegraphics[width=0.95\linewidth]{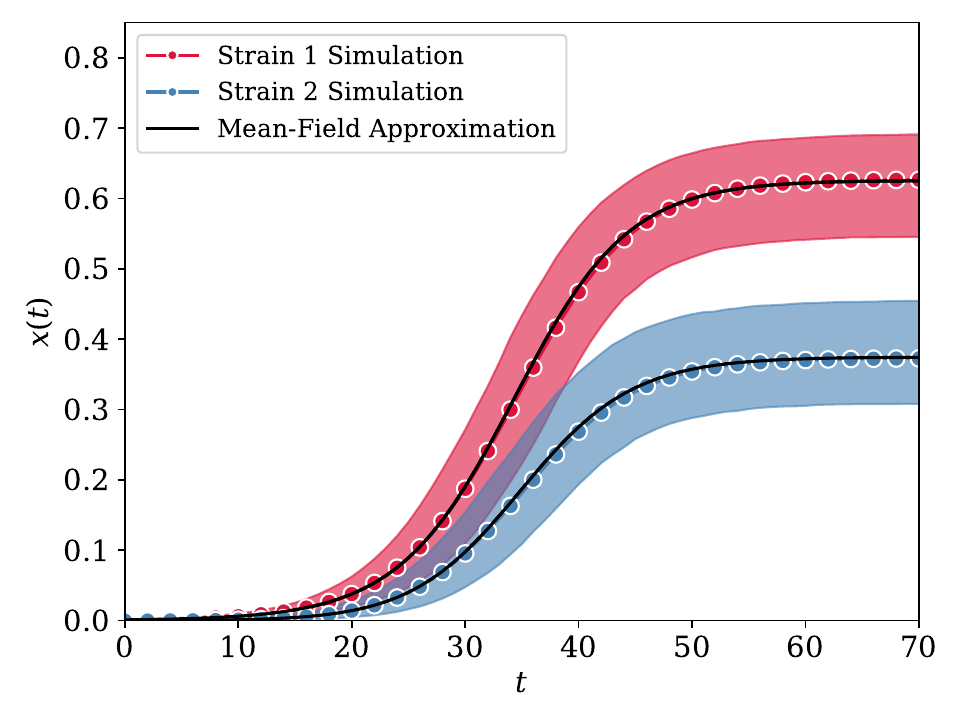}
    \caption{\textbf{Validation of the well-mixed mean-field approximation for the Noisy SI model with binary symmetric channel at $\epsilon = 0.01$}. Dotted curves: stochastic simulation averages over 1000 realizations ($N = 10^4, \beta = 0.05, \meank = 4, \bm{x}(0) = (10/N, 0)$), shaded regions indicate $95\%$ confidence intervals. Solid black curves: numerical integration of~\eqref{eq:NSI_mf}. Agreement is excellent across the full trajectory.}
    \label{fig:mixed_mf_fit}
\end{figure}

Figure~\ref{fig:mixed_mf_fit} demonstrates the agreement between the mean-field approximation of Eq.~\eqref{eq:NSI_mf} and stochastic simulations of the underlying reaction system~\eqref{eq:general_reaction} in a well-mixed population. 

Defining the transmission kernel as the binary symmetric channel, 
\[
Q_\mathrm{BS} = \left(
\begin{array}{cc}
    1 - \epsilon & \epsilon \\
    \epsilon & 1 - \epsilon
\end{array}
\right),
\]
with $\epsilon=0.01$, the mean-field trajectories of both strains lie within the $2\sigma$ confidence band of the simulation averages across the full dynamical range. The endpoint of each strain matches the prediction of Eq.~\eqref{eq:stable_mf} to within sampling noise, providing a direct check that the exact solution derived above describes the long-time behavior of the stochastic process.

\begin{figure*}[t!]
    \centering
    \includegraphics[width=.9\textwidth]{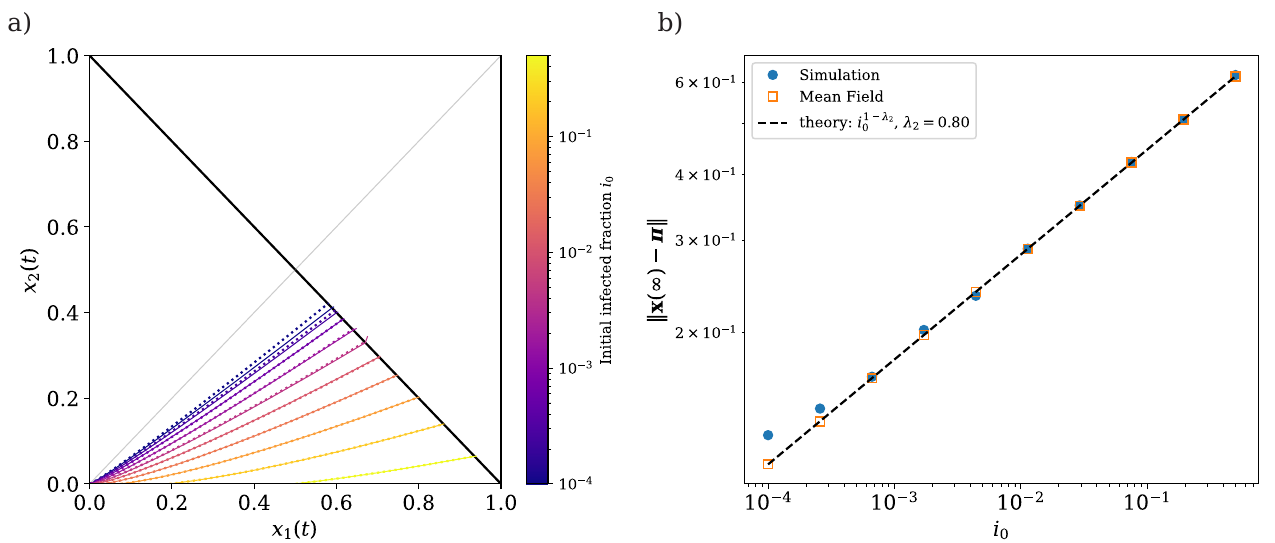}
    \caption{\textbf{Mean-field and simulation agree on how $i_0$ and $Q$ shape dynamics.} \textbf{a)} Trajectories in the $(i_{1},i_{2})$-plane for a fixed, column-stochastic $Q_\mathrm{BS}$ with $\epsilon=0.1$. Trajectories are seeded at $(i_{0},0)$ (i.e. all initial mass in strain $1$) for $i_{0}$ varying logarithmically over a chosen range, and terminate on the diagonal $i_{1}+i_{2}=1$ as required by the SI constraint. \textbf{2} Solid curves show stochastic simulation averages over 1000 realizations; dashed curves show integrations of the mean-field system. The dominant right eigenvector $\bm{\pi}$ is marked on the diagonal. Terminal points slide monotonically from the initial-strain vertex toward $\bm{\pi}$ as $i_{0}$ decreases, in quantitative agreement with Eq.~\eqref{eq:spectral-attractor}.}
    \label{fig:convergence_to_pi}
\end{figure*}

\section{Spectral Decomposition of the Attractor}\label{sec:spectral}

The exact solution of the well-mixed dynamics reduces the long-time behavior to a single matrix
exponential governed by $Q$. In this section
we exploit this structure to characterize the attractor
$\bm{x}(\infty)$ and the manner in which it is approached, and show that both are determined by the spectrum of $Q$.

\subsection{Eigenvalue Expansion of the Attractor}

To understand how the $Q$ governs the long-time dynamics of the system, we begin by considering all $Q$ that are irreducible, aperiodic, and diagonalizable. We henceforth only consider such $Q$, except where explicitly stated. Column-stochasticity entails $\lambda_{1}=1$ is an eigenvalue of $Q$ with left eigenvector
$\bm{v}_{1} = \bm{1}$ and right eigenvector
$\bm{u}_{1} = \bm{\pi}$. Thus, if interpreted as the transition matrix of a Markov chain, $\bm{\pi}$ is the stationary distribution of $Q$. To illuminate the relationship between the Noisy SI model and linear transmissions characterized by a random walk on $Q$, we hereafter refer to $\bm{\pi}$ as the stationary distribution. We also normalize so that $\bm{1}^{\top}\bm{\pi} = 1$. All
remaining eigenvalues satisfy $|\lambda_{m}| < 1$, thus the spectral gap, $1-|\lambda_{2}|$, is strictly positive.

Using the biorthogonal decomposition
\[
    Q = \sum_{m} \lambda_{m}\, \bm{u}_{m}\bm{v}_{m}^{\top}
\]
with $\bm{v}_{j}^{\top}\bm{u}_{m} = \delta_{jm}$, we find 
\[
    \exp\!\big(\beta \meank cY(\infty)\, Q\big)
    = \sum_{m} e^{\beta \meank cY(\infty)\lambda_{m}}\,
      \bm{u}_{m}\bm{v}_{m}^{\top}.
\]
Writing $\bm{x}(0) = i_{0}\tilde{\bm{x}}_{0}$, where
$\tilde{\bm{x}}_{0}$ is the probability vector specifying the composition of the initial seed across infected strains and $i_0 = I/N$ at time $t=0$, the $m=1$
term collapses exactly to $\bm{\pi}$, as $\bm{v}_{1}^{\top}\bm{x}(0) = i_{0}$ and
$e^{\beta \meank cY(\infty)} = 1/i_{0}$. The remaining terms form a spectral
expansion in the sub-leading right eigenvectors of $Q$,
\begin{equation}
    \bm{x}(\infty)
    = \bm{\pi}
    + \sum_{m\geq 2}
      i_{0}^{\,1-\lambda_{m}}\,
      (\bm{v}_{m}^{\top}\tilde{\bm{x}}_{0})\,
      \bm{u}_{m}.
    \label{eq:spectral-attractor}
\end{equation}
Equation~\eqref{eq:spectral-attractor} is the principal structural
result of this section. The attractor decomposes into a universal
component $\bm{\pi}$---the stationary state of a random walk with transition matrix $Q$,  independent of the initial composition-- and perturbative corrections along each sub-leading eigendirection, suppressed by the factor $i_{0}^{1-\lambda_{m}}$.

\subsection{Geometric Interpretation}

For each value of $i_0$, Eq.~\eqref{eq:spectral-attractor} defines a simplex of possible attractor states inside the larger $(|\mathcal{A}|-1)$-simplex of strain compositions. These simplices are nested such as that, as $i_0$ decreases, the corresponding simplex contracts toward $bm{\pi}$. For defective $Q$, the image may collapse to a lower-dimensional face
of the simplex, but we restrict attention to diagonalizable $Q$. In
these settings, two limits are immediate. As $i_{0} \to 0$, every
correction vanishes and the image of the initial simplex contracts to
the single point $\bm{\pi}$, meaning that rare seeding erases all
memory of the initial composition. On the other hand, somewhat
trivially, as $i_{0} \to 1$, then $Y(\infty) \to 0$ and the matrix
exponential approaches the identity, so $\bm{x}(\infty) = \bm{x}(0)$
and no spreading occurs. Thus, the size of the seed tunes the extent to which the spreading dynamics can act. Dense seeding leaves the population frozen near its initial composition while sparse
seeding allows for more transmission events and thus more mixing, driving the system to a steady-state which forgets the initial condition and is determined only by $Q$. This can be seen in Figure~\ref{fig:convergence_to_pi}a, which shows how trajectories fixed as $i_- = (i_1, 0)$ vary with $i_i$ for a fixed $Q$. 

For intermediate $i_{0}$, the map $\tilde{\bm{x}}_{0} \mapsto
\bm{x}(\infty)$ is affine, and the image of the initial simplex is a
contracted copy nested around $\bm{\pi}$. The contraction is
generically anisotropic so that along the eigendirection $\bm{u}_{m}$, the
image is compressed by the factor $i_{0}^{1-\lambda_{m}}$. Modes with
$\lambda_{m}$ close to 1 retain visible memory of the initial
condition, while modes with small $\lambda_{m}$ decouple rapidly. The
attractor family thus traces a nested sequence of simplices whose
shape is fixed by the spectrum of $Q$ and whose overall scale is set
by $i_{0}$. Altogether, it is therefore $i_0$ and the spectrum of $Q$ which control the long-time behavior of the system. In the following sections, we treat these two dependencies in turn.

\paragraph{Dependence on $i_0$}
As shown in Figure~\ref{fig:convergence_to_pi}, for a fixed initial distribution $\tilde{\bm{x}}$, the displacement of the attractor from $\bm{\pi}$ shrinks as $i_0$ decreases, at a rate set by the spectral gap of $Q$. Concretely, the correction in Eq.~\eqref{eq:spectral-attractor}
is driven by $i_0^{1-\lambda_2}$, as higher-mode contributions decay
strictly faster in the limit $i_0 \to 0$. This renders the asymptotic
scaling
\begin{equation}\label{eq:asymptotic-gap}
    \lVert \bm{x}(\infty) - \bm{\pi} \rVert
    \;\sim\;
    i_{0}^{\,1-\mathrm{Re}\,\lambda_{2}}
    \qquad (i_{0}\to 0),
\end{equation}
wherever $\tilde{\bm{x}}_0 \neq \bm{\pi}$. As shown in Figure~\ref{fig:convergence_to_pi}b, this means that the displacement of the steady state of the dynamics from the stationary distribution of $Q$ mapped against $i_0$ has slope $1 - \mathrm{Re}\,\lambda_2$,
allowing the spectral gap of the transmission kernel to be inferred
from endpoint statistics of simulated or observed spreading
dynamics without directly diagonalizing $Q$.

\begin{figure*}[t]
    \centering
    \includegraphics[width=0.9\textwidth]{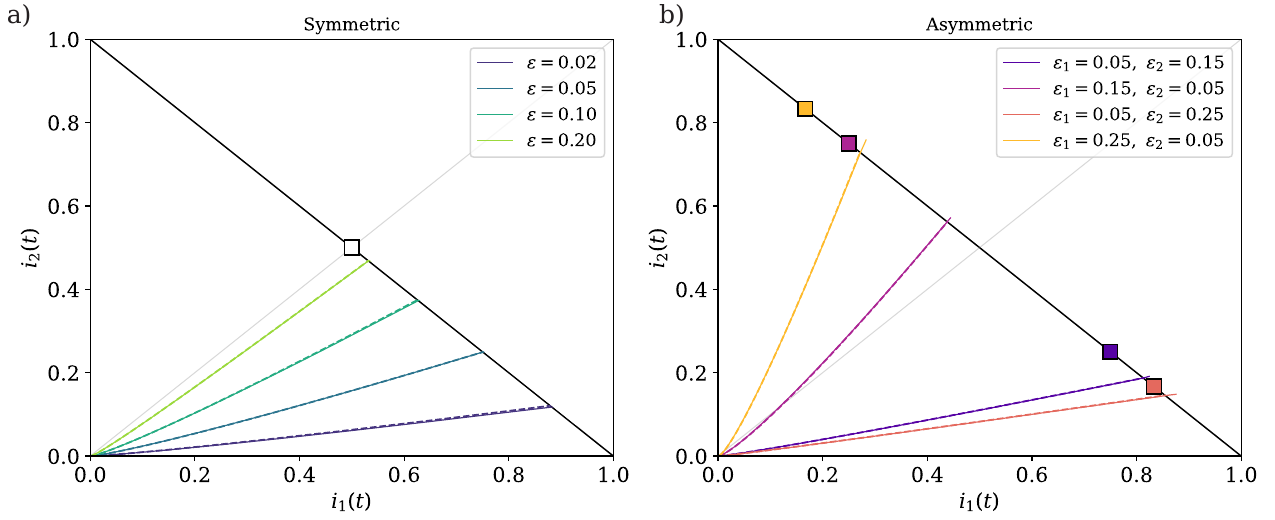}
    \caption{\textbf{Channel asymmetry separately controls the location and the contraction rate of the attractor in the well-mixed two-strain model.} Solid curves show the stochastic simulation averages over 1000 realizations. Dashed curves show the numerical integration of Eq.~\eqref{diffeq_vec}. All trajectories are seeded with $i_0 = 10^{-3}$ all on $\stateI_1$ and all other parameters are as described in Figure~\ref{fig:mixed_mf_fit}. \textbf{a)} Trajectories for symmetric $Q$. Whenever $Q$ is symmetric, the stationary distribution $\bm{\pi}$ (shown as a white square) is the uniform distribution in $|\mathcal{A}|$ dimensions. \textbf{b)} Trajectories for asymmetric $Q$ with their corresponding stationary distributions (shown as colored stars).}
    \label{fig:varying_q}
\end{figure*}

\paragraph{Dependence on the structure of $Q$}
Equation~\eqref{eq:spectral-attractor} implies that the entire
attractor family is fixed by the spectrum of $Q$. Parameterizing $Q$
as 
\[
    Q = \left(\begin{array}{cc}
       1-\epsilon_1  & \epsilon_2 \\
       \epsilon_1  & 1-\epsilon_2
    \end{array}\right)
\]
by its noise rates, $\epsilon_1, \epsilon_2$, we can observe how the
attractor geometry is systematically deformed. Solving $Q\bm{\pi} =
\bm{\pi}$, we get the form 
\begin{equation}\label{eq:pi_calc}
    \bm{\pi} = \frac{1}{\epsilon_1 + \epsilon_2} \left(\begin{array}{c}
        \epsilon_2 \\
        \epsilon_1 
    \end{array} \right)
\end{equation}
showing explicitly that the ratio of strain abundances at $\bm{\pi}$ is set by the inverse ratio of mutation rates. We can use this expression to
consider two limits. First, the noiseless channel $Q = I$ is
degenerate---every eigenvalue is 1, the contraction factor is
$(i_{0})^{0} = 1$, and the attractor simplex coincides with the
initial simplex. This, therefore, recovers the classical multi-strain
SI model in which no mixing between strains occurs. Meanwhile, the
maximally noisy channel in which $\epsilon_1 = \epsilon_2 = ... = \epsilon_{|\mathcal{A}|} = 1/|\mathcal{A}|$ has
$\lambda_{m} = 0$ for $m\geq 2$. Thus, the contraction factor reduces
to $i_{0}$, and the attractor collapses almost immediately onto
$\bm{\pi}$. As shown in Figure~\ref{fig:varying_q}, the
stationary state for all symmetric $Q$ is the uniform distribution in
$|\mathcal{A}|$ dimensions and as $\epsilon$ increases, we see the
steady-state of the dynamics approach this value. Meanwhile, when
$\epsilon_j \neq \epsilon_k$ for any $j,k \in \mathcal{A}$, the stationary distribution shifts as
\eqref{eq:pi_calc} and the dynamics are pulled towards this
distribution proportionally with $\lambda_2$. In summary, then, we might think of the two structural roles of the channel matrix, Q, as being cleanly separated. Its eigenvector, $\bm{\pi}$, sets where the dynamics go, while its spectral gap sets how forcefully they are pulled there.

\section{Pairwise Mean Field}\label{sec:structured}

Relaxing the assumption of a fully connected population to instead consider structured populations, the dynamics of the model can be approximated using a pairwise approximation \cite{kiss_mathematics_2017}. This approach considers the expected number of individuals infected as a function of the expected number of edge-linked \textit{pairs} of type $\stateS \stateI_k$ (one susceptible and one $k$-infected individual which share an edge) at a given moment in time. These models are closed by assuming an approximation for the number of triples of type $\stateI_j \stateS \stateI_k$ and $\stateS \stateS \stateI_j$ based on the population topology. 

In this approximation, the expected change in the number of nodes in state $\stateI_j$ at any given time is given as 
\begin{equation}\label{eq:full_pairwise_di}
    \ddt [ I_j ] = \sum_{k \in \mathcal{A}} \beta Q_{jk} [SI_k]
\end{equation}
where $[SI_k]$ is the expected number of edges with a susceptible node at one end and a node in state $\stateI_k$ at the other. 

Ascertaining the dynamics of edge states becomes more complex. To begin, we first recognize that the expected influx of $\stateS \stateI_j$ edges at each time step comes from changes in triples (i.e. triangles and cherries). An $\stateS \stateI_j$ edge is thus formed when a successful $j \to j$ infection event occurs in an $\stateS \stateS \stateI_j$ triple, or a $k \to j$ mutation occurs in an $\stateS \stateS \stateI_k$ triple. 

We decompose these parts as $f^\mathrm{in}$, $f^{out, 2}$, $f^\mathrm{out,3}$ so that 
\begin{equation}\label{eq:fall}
    \ddt [ SI_j ] = f^\mathrm{in} + f^\mathrm{out,2} + f^\mathrm{out,3} 
\end{equation}

\noindent We can thus capture the expected influx of $\stateS \stateI_j$ edges as 

\begin{equation}\label{eq:fin}
   f^\mathrm{in}  = \sum_k^\mathcal{A} \beta Q_{jk} [ SSI_k ],
\end{equation}

\noindent the outflux from infections occurring in $\stateS\stateI_j$ edges as 
\begin{equation}\label{eq:f2}
   f^\mathrm{out,2}  = - \sum_k^\mathcal{A} \beta Q_{kj} [ SI_j ],
\end{equation}

\noindent and the outflux from infections occurring in $I_kSI_j$ edges as
\begin{equation}\label{eq:f3}
    f^\mathrm{out,3} = - \sum^\mathcal{A}_{k}\left(\sum^\mathcal{A}_l \beta Q_{lk}\right)[ I_kSI_j ]. 
\end{equation}

% meaning f_in 
In $f^\mathrm{in}$ for strain $j$ we are counting all the $\stateS \stateS \stateI_k$ triples in which the infected node infects an $ \stateS$ to produce an $ \stateI_j$.
% meaning f_out_2
In $f^{out, 2}$ for strain $j$ we are counting all the $ \stateS \stateI_j$ edges lost due to infection of the $\stateS$ node by the $\stateI_j$ node to any strain (including $j$). 
% meaning f_out_3
In $f^\mathrm{out,3}$ we count all the $\stateS \stateI_j$ edges that are part of $\stateI_k \stateS \stateI_j$ triples ($\forall k \in \mathcal{A}$ including $j$) which are converted to $\stateI_k \stateI_\ell \stateI_j$ ($\forall \ell \in \mathcal{A}$) via infection from a node ``outside'' of the $\stateS \stateI_j$ edge of interest. 

Finally, we must track the number of edges in which nodes at both ends are still susceptible. Because the model has only infections and no recovery, these edges can only be destroyed. Specifically, they are destroyed with any infection event. As such, we have the expected number of $\stateS \stateS$ edges changes as 
\begin{equation}\label{eq:ss}
    \ddt [ SS ] = - 2\sum_k^\mathcal{A} \sum_l^\mathcal{A}  \beta Q_{lk} [ SSI_k ] 
\end{equation}
which depicts the loss of an $\stateS \stateS$ edge occurring when the infected node in an $\stateS \stateS \stateI_k$ triple infects either of the two susceptible nodes (giving us the factor of two in front).

As $f^\mathrm{in}$, $f^\mathrm{out,3}$, and $\stateS \stateS$ all involve third moments, we employ the following moment closure approximation from~\cite{kiss_mathematics_2017} for arbitrary compartments $A, B$:

\begin{equation}\label{eq:closure}
    [ ASB ] \approx \kappa \frac{[ AS ] [ S B ]}{ [ S ]},
\end{equation}
where, from~\cite{kiss_necessary_2022}, $\kappa = \frac{\bar{k}-1}{\bar{k}}$ for $\bar{k}$-regular graphs while $\kappa = 1$ for Poisson-distributed graphs. This approximation makes the simplifying assumption that susceptible and infected nodes are randomly distributed across the network. 

Putting together~\eqref{eq:fall}-\eqref{eq:closure}, we obtain 
\begin{widetext}
\begin{align}
    \ddt [SI_j] &= \frac{(\meank -1)}{\meank[S]}\left([SS]\sum_{k \in \mathcal{A}} \beta Q_{jk}[SI_k] - \sum_{k \in \mathcal{A}}(\sum_{\ell \in \mathcal{A}} \beta Q_{\ell k} [SI_j][SI_k]\right) - \sum_{k \in \mathcal{A}} \beta Q_{k j}[SI_j] \label{eq:full_pairwise_si}\\
    \ddt[SS] &= \frac{2[SS](1-\meank)}{\meank[S]} \sum_{k \in \mathcal{A}}\sum_{\ell \in \mathcal{A}} \beta Q_{\ell k}[SI_k], \label{eq:full_pairwise_ss}
\end{align}
which, alongside~\eqref{eq:closed_S} and~\eqref{eq:full_pairwise_di}, constitute the pairwise approximation of a $\meank$-regular graph. 
\end{widetext}

\begin{figure}
    \centering
    \includegraphics[width=\linewidth]{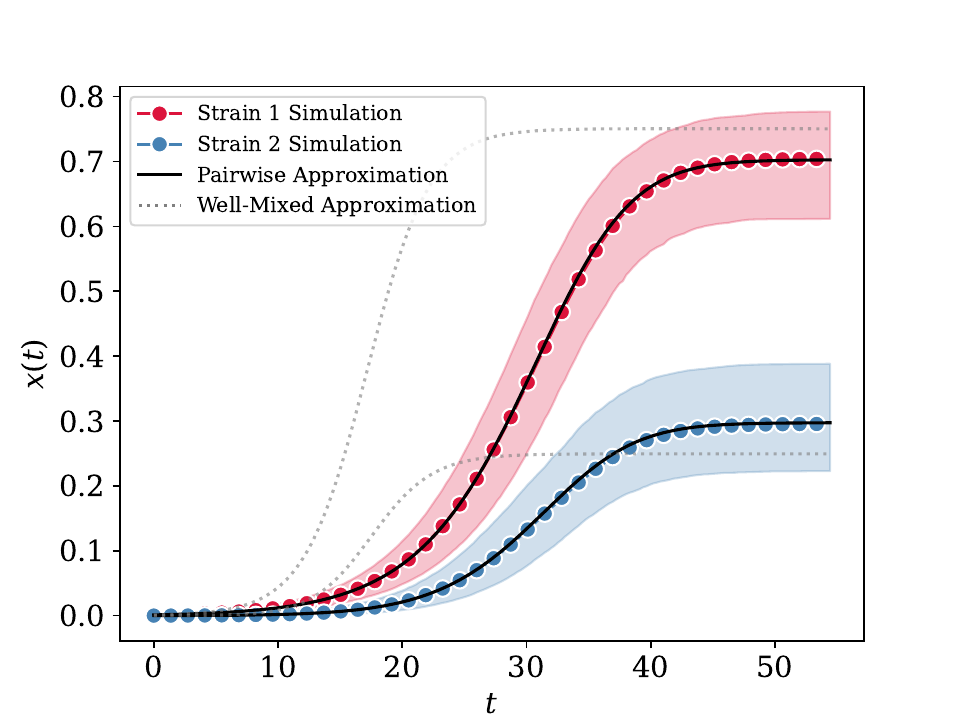}
    \caption{\textbf{Validation of the pairwise mean-field approximation on a $\bar{k}$-regular graph.} Dotted curves show stochastic simulation averages with $5$th--$95$th percentile confidence intervals over $1000$ realizations on a random $\bar{k}$-regular graph ($N = 10^4$, $\meank =4$, $\beta=0.05$, $\epsilon = 0.05$, $i_0 = 10/N$ all in strain~$1$). Black curves show numerical integration of the pairwise system of Sec.~\ref{sec:structured}. Dotted grey lines show the well-mixed mean-field approximation for the same initial conditions. Agreement between the pairwise approximation and simulations is excellent through the rapid spreading phase, with some simulation lag during the plateau phase.}
    \label{fig:regular_graph_fit}
\end{figure}

\paragraph{Pairwise dynamics on a regular graph}
Figure~\ref{fig:regular_graph_fit} shows the agreement between stochastic simulations of the Noisy SI process on a $\meank$-regular graph and numerical integration of this system for a dynamics with two strains. The pairwise prediction lies within the confidence band of the simulation across the full trajectory. Two qualitative differences from the well-mixed dynamics are visible
in the figure. First, as expected, the onset of spreading is delayed in every pairwise approximation compared to the well-mixed approximation because assumptions about $\stateS \stateI$ edges in the latter create an artificially faster early growth rate \cite{pastor-satorras_epidemic_2015}. More relevant to the multi-strain dynamics, however, we also find that the saturation plateau is shifted, as both strains approach long-time values closer to the spectral stationary distribution $\bm{\pi}$ than in the well-mixed case, despite identical $Q$. 

\begin{figure}
    \centering
    \includegraphics[width=\linewidth]{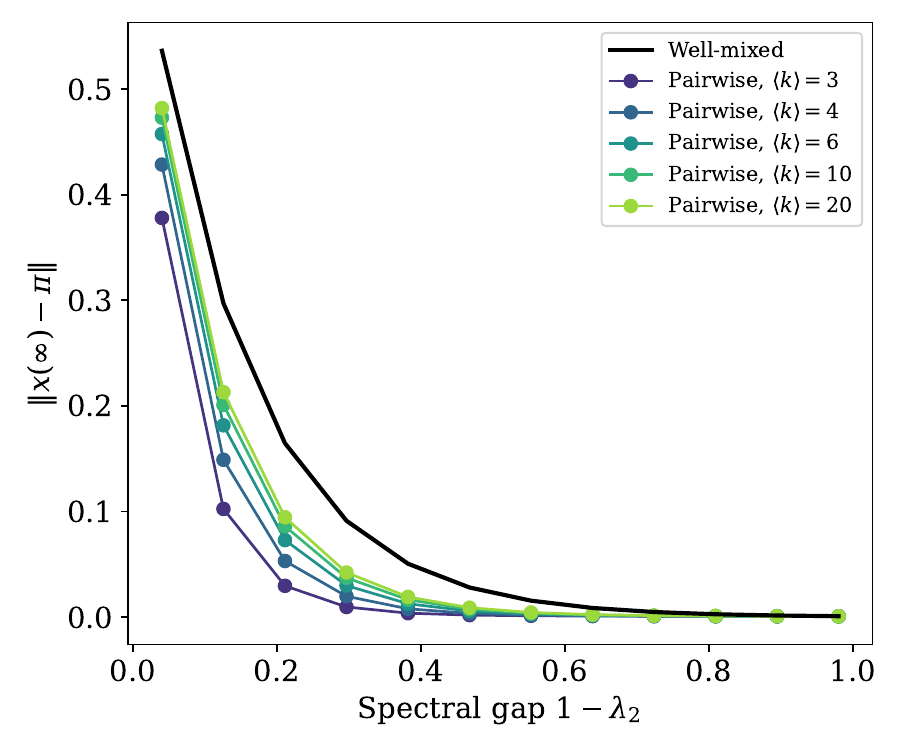}
    \caption{\textbf{Network sparsity drives the pairwise attractor toward the spectral stationary distribution.} Distance from $\bm{\pi}$, the attractor defined by the spectrum of $Q$, for the well-mixed approximation (black) and the pairwise approximation on $\meank$-regular graphs at several values of $\meank$ (colored), as a function of the spectral gap $1-\lambda_2$ of the binary symmetric channel, $Q_\mathrm{BS}$. All values were obtained by integrating the corresponding mean-field systems to saturation, with $i_0 = 10^{-3}$ all beginning in the first strain.}
    \label{fig:graph_gap}
\end{figure}

This is further examined in Figure~\ref{fig:graph_gap}, which shows the systematic relationship between graph degree and the location of the attractor predicted by the pairwise approximation. For all $\bar{k}$ tested, the pairwise attractor sits closer to the spectral stationary distribution $\bm{\pi}$ than the well-mixed attractor, with the effect most pronounced at intermediate spectral gaps and small $\bar{k}$. Heuristically, the pairwise closure encodes the slower per-node spreading rate, extending the effective time available for the channel $Q$ to mix strain identities per unit of forward propagation. The result is that network sparsity acts on the long-time behavior in a manner qualitatively similar to an increase in the channel noise rate---in both cases, the additional mixing pulls the attractor closer to $\bm{\pi}$.

\section{Communication Through a Population}\label{sec:communication}

A natural reinterpretation of this model casts it as a model of communication across a population. In this register, the strains of $\mathcal{A}$ are not biological variants but discrete signals or messages, and the channel $Q$ encodes the per-transmission fidelity of message reproduction. As an example, one might imagine the telephone game—a game in which the first speaker produces an utterance that is increasingly distorted as it propagates through a chain of subsequent speakers. The propagation of strains through the population then corresponds to the spread of these progressively distorted utterances, each a mutated descendant of the initial signal. More generally, the model is then interpreted as describing the spread of a noisy signal from an initially informed subset to the entire population, with each peer-to-peer transmission introducing fixed-rate distortion.

\paragraph{Setup} Let $W$ denote a source that can take one of $\Omega$ discrete states
with distribution $p(\omega)$. At time $t=0$, a small
fraction $i_0$ of the population observes the source and encodes its
state through a map $\phi: \Omega \to \mathcal{A}$.  The remaining
population is uninformed and learns about $W$ only through
peer-to-peer transmission governed by the Noisy SI dynamics of
Sec.~\ref{sec:model}. The composition, $\tilde{\bm{x}}_0$, of the
initial informed subset is then determined by the joint distribution
of the source, $p(\omega)$, and the encoding, $\phi$.

This setup formalizes a class of scenarios in which information about
a localized or hard-to-access source must propagate through a population by
serialized, noisy communication. Examples include rumor cascades originating from a
small number of eyewitnesses, news propagating from journalists
through informal networks of readers, or scientific findings moving
from primary literature into popular understanding. In each case one is not interested in the quantity of any particular strain, but how the distribution of strains at initial encoding and throughout the cascade preserves source information.

\subsection{Applications in Information Spreading}

\paragraph{Mutual Information} A natural observable emerging from this framing is the mutual information
\[
I[W; x_t] = \sum_{\omega \in \Omega}\sum_{k \in \mathcal{A}} p(\omega)\tilde{x}_k(t) \log \frac{\tilde{x}_k(t)}{\sum_{\omega^\prime \in \Omega} p(\omega^\prime) \tilde{x}_k(t)} \label{eq:MI_phi} 
\]
between the source and the distribution of messages at time $t$, which quantifies how much
about $W$ can be inferred from sampling the population at time, $t$. 

As a functional of the trajectory $\bm{x}(t)$, the long-term behavior of this quantity is also predominantly determined by the stationary distribution
$\bm{\pi}$ of the channel $Q$. This implies that the long-time
mutual information $I[W; x(\infty)]$ is bounded above by a function of
$i_0$ and the spectrum of $Q$, and approaches zero as $i_0 \to 0$
regardless of how informative the encoding $f$ is at the source.
Information about $W$ is degraded by the same spectral mechanism that
contracts the simplex of strain compositions toward $\bm{\pi}$.

\paragraph{Implications for Information Spreading}
This perspective makes several qualitative predictions that distinguish the present framework from simpler models of information
spread. First, the asymptotic information content of the cascade depends on the source distribution and the encoding only through their projection onto the sub-leading eigenspace of $Q$. Encodings
that exploit the slowly-mixing directions of $Q$ (i.e., those for which
$\phi(\omega)$ varies along eigenvectors with $\lambda_m$ close to
$1$) retain more information than encodings aligned with rapidly
mixing directions, even at fixed $i_0$.

Second, comparisons between the well-mixed and pairwise
approximations in Sec.~\ref{sec:structured} acquires a communicative interpretation. In this light, we can read the results as saying that sparse contact networks dilute information about $W$ more aggressively than dense ones, because the slower per-node spreading rate gives the channel more opportunity to mix message identities before saturation. Thus, the location of the long-time attractor is shifted toward  $\bm{\pi}$, which carries no information about the source.

As a full information-theoretic treatment lies beyond the scope of this work, the aim of this section is to indicate that the analytical structure developed in Secs.~\ref{sec:wellmixed}- \ref{sec:structured} carries directly into the communicative setting. Morover, it is to show that the spectral characterization of the attractor governs the long-time fidelity of population-level communication in the same way that it governs the strain composition of an epidemic.

\section{Discussion}\label{sec:discussion}

The Noisy SI model studied in this work occupies a deliberately
minimal position in the multi-strain compartmental literature. By
dropping cross-immunity, competition, recovery, and structural
constraints on the mutation kernel, what remains is a model whose
mean-field dynamics admit an exact closed-form solution and whose
long-time behavior is characterized entirely by the spectrum of a
single matrix. The principal contributions of this work---the exact
solution in the well-mixed regime, the spectral form of the
attractor, and the pairwise extension to structured populations---are
analytical results that hold for arbitrary column-stochastic $Q$ and
that we have validated against stochastic simulations.

The simplicity of the model is both a strength and a limitation. Real epidemic processes typically involve recovery, immunity, and rich strain interactions that the present framework abstracts away. These abstractions may nevertheless be reasonable in regimes where spreading is fast relative to recovery or immune response and where strain interactions are weak, as well as in non-epidemiological applications such as the communicative setting developed in Sec.~\ref{sec:communication}. A separate caveat applies to the mean-field approximations themselves, which hold cleanly only on fully connected populations and on $\meank$-regular graphs. Extensions to heterogeneous-degree, modular, or temporally varying contact structures would require correspondingly extended closure schemes. We expect the spectral characterization of Sec.~\ref{sec:spectral} to remain qualitatively intact in these extensions, since the location of the attractor is determined by $Q$ alone, but the rate of convergence and the magnitude of the sparsity-induced shift identified in Sec.~\ref{sec:structured} will likely acquire topology-dependent corrections.

Finally, our communicative reinterpretation presented in
Sec.~\ref{sec:communication} makes available comparisons with a
litany of models and empirical findings concerning the social
transmission of heterogeneous content. For instance, iterated learning experiments
have shown that transmission chains converge to systematic
distortions of input signals reflecting biases of the transmitting
agents~\cite{kirby_iterated_2014} and studies of online information spreading have documented that different categories of content exhibit markedly different propagation dynamics on the same underlying network~\cite{romero_differences_2011, goel_structural_2015, berger_what_2012}. Perhaps the most salient analogy comes from the Paris School tradition in cultural
evolution, which has framed the convergence of representations across a population in terms of ``cultural attractors''~\cite{claidiere_cultural_2014}. So, while our model is considerably simpler than any that may be found in these literatures, it shares the basic observation that the long-time composition of messages in a population is shaped in part by the noise and mutation inherent to transmission, rather than the original signal alone. The closed-form solution and spectral characterization developed here may thus be useful for studying this kind of convergence in a setting where the underlying dynamics are fully analytically accessible.

\section{Acknowledgements}
The authors would like to thank P\'eter L. Simon of E\"otv\"os Lor\'and University for his invaluable contributions to solving the well-mixed mean-field approximation. 

%\appendix

%\bibliography{references}% Produces the bibliography via BibTeX.

\end{document}